\documentclass[aip,apm,reprint]{revtex4-1}

\usepackage[breaklinks=true,colorlinks=true,linkcolor=blue,urlcolor=blue,citecolor=blue]{hyperref}
\usepackage{graphicx}
\usepackage{dcolumn}
\usepackage{bm}

\usepackage[utf8]{inputenc}
\usepackage[T1]{fontenc}
\usepackage{mathptmx}
\usepackage{etoolbox}

\usepackage{siunitx}
\usepackage{upgreek}

\makeatletter
\def\@email#1#2{%
	\endgroup
	\patchcmd{\titleblock@produce}
	{\frontmatter@RRAPformat}
	{\frontmatter@RRAPformat{\produce@RRAP{*#1\href{mailto:#2}{#2}}}\frontmatter@RRAPformat}
	{}{}
}%
\newcommand*{\rom}[1]{\expandafter\@slowromancap\romannumeral #1@}
\makeatother

\begin{document}

\title{Spin-wave emission with current-controlled frequency by a PMA-based spin-Hall oscillator}

\author{M. Bechberger}%
\altaffiliation{M. Bechberger and D. Breitbach contributed equally to this work.}
\email{mbechber@rptu.de}
\author{D. Breitbach}%
\altaffiliation{M. Bechberger and D. Breitbach contributed equally to this work.}
\author{A. Koujok}
\author{B. Heinz}
\affiliation{Fachbereich Physik and Landesforschungszentrum OPTIMAS, \\Rheinland-Pfälzische Technische Universit{\"a}t Kaiserslautern-Landau, \mbox{D-67663 Kaiserslautern, Germany}}
\author{C. Dubs}
\affiliation{\mbox{INNOVENT e.V. Technologieentwicklung, D-07745 Jena, Germany}}
\author{\mbox{A. Hamadeh}}
\affiliation{\mbox{Université Paris-Saclay, Centre de Nanosciences et de Nanotechnologies, CNRS, 91120, Palaiseau, France
}}

\author{\mbox{P. Pirro}}
\affiliation{Fachbereich Physik and Landesforschungszentrum OPTIMAS, \\Rheinland-Pfälzische Technische Universit{\"a}t Kaiserslautern-Landau, \mbox{D-67663 Kaiserslautern, Germany}}

\date{\today}

\begin{abstract}
Spin-torque and spin-Hall oscillators (SHOs) have emerged as promising candidates for building blocks in neuromorphic computing due to their ability to synchronize mutually,  a process that can be mediated by propagating spin waves. 
We demonstrate a SHO that takes advantage of a low-damping magnetic garnet with dominant perpendicular magnetic anisotropy (PMA), namely gallium-substituted
yttrium-iron-garnet (Ga:YIG). In-plane magnetized Ga:YIG allows for the operation at a high efficiency level while also enabling resonant spin-wave emission. A nonlinear self-localization of the excitation is avoided by exploiting the positive nonlinear frequency shift, which facilitates a current-controlled frequency of the emitted spin waves. 
Via micro-focused Brillouin light scattering spectroscopy, we investigate the properties of the local auto-oscillation and its spin-wave emission. Multiple modes are excited and compete internally, with two propagating modes detected up to distances larger than \SI{10}{\micro\meter}. Their frequencies combine to an extended frequency bandwidth of approximately \SI{1.6}{\giga\hertz}. 
The experimentally observed two-mode system and its transition to a single mode at higher currents are reproduced via micromagnetic simulations, which account for spatial variation of the PMA arising due to the microstructures on Ga:YIG.
Our results propose a promising platform for hosting SHOs, interconnected via propagating spin waves, with particular relevance to neuromorphic computing.
\end{abstract}

\pacs{}
\maketitle

\section{Introduction}
The field of neuromorphic computing is inspired by the human brain, a highly interconnected network of nonlinear neurons, which achieves unprecedented energy efficiency for information processing. Consequently, the development of nanoscaled elements for hardware-based neuromorphic approaches has become subject of significant interest\cite{ross.neural-network-multilayer.2023,rodrigues.neural-network-stno.2023}. 
In this regard, spin-torque oscillators and spin-Hall oscillators (SHOs) are of particular interest in the field of spintronics\cite{demidov.reportSHNO.2017,li.reportSHNO.2020,grollier.neuromorphic.2020}. The two approaches rely on the application of spin-orbit torques (SOT), such as the spin-transfer torque \cite{slonczewski.STT.1996,slonczewski.STT.1999,ralph.STT.2008}, compensating the natural damping of a magnetic material to achieve coherent auto-oscillations \cite{slavin.STNO.2009}. In particular, their ability to synchronize to external sources \cite{rippard.synchronization-external.2005, demidov.synchronization-external.2014} or to interact mutually \cite{kaka.synchronization.2005} is a key feature for their implementation in neural networks. While short-range coupling can be mediated through dipolar or exchange interactions \cite{locatelli.STNO-vortex.2015,kurokawa.STNO-exchange.2022}, long-range coupling has been demonstrated utilizing electric currents, requiring energy-inefficient conversions between magnetic and electric domains \cite{grollier.synchronization.2006,lebrun.synchronization.2017}.
Therefore, using spin waves to couple through the magnetic medium itself emerges as a promising approach.
A key requirement for this coupling is the ability of the auto-oscillation to resonantly emit spin waves into the surrounding spin-wave medium. This is only possible if the auto-oscillation frequency within the spin-injection region corresponds to a propagating mode in the surrounding area. 
It has been demonstrated that inducing perpendicular magnetic anisotropy (PMA) is effective for SHOs to operate within the positive nonlinear frequency shift regime, which appears to enable spin-wave emission\cite{fulara.emission.2019, succar.PMA-emission.2023}. Nevertheless, those systems often require the magnetization oriented at an oblique out-of-plane angle, reducing the efficiency of the spin Hall effect\cite{hirsch.SHE.1999}(SHE). Accordingly, the (anti-)damping torque is maximized for in-plane magnetized SHOs\cite{divinskiy.nanoconstriction.2017}.
In such systems based on metallic ferromagnets, spin-wave mediated synchronization has already been achieved, but the devices are limited to short distances due to high spin-wave damping \cite{kaka.synchronization.2005,zahedinejad_2D-synchronization_2020, kumar.synchronization.2025}. Consequently, low-damping materials such as yttrium iron garnet (YIG) offer great potential for longer interaction ranges \cite{dubs.LPE.2020,heinz.propagation.2020}. Spatially localized magnetization dynamics have been observed in YIG/Pt-based SHOs\cite{collet.yig-pt.2016,demidov.yig-pt.2016,schneider.boseeinstein.2020,schneider.stabilization.2021} due to the negative nonlinear frequency shift for the in-plane magnetized configuration. In this case, spin-wave emission can still be enabled by exploiting the band structure, achieving nonlinear scattering to other spin-wave modes, such as PSSWs \cite{schlitz.sw-emission.2025}.
Another approach has been demonstrated in Bi-substituted YIG\cite{soumah.biyig.2018}, whose PMA was chosen to eliminate the nonlinear frequency shift, leading to the emission of coherent spin waves with notable propagation distances\cite{evelt.biyig.2018}. However, both approaches are limited by the absence of frequency tunability, which can be crucial for neuromorphic computing\cite{romera.vowel.2018}.\\
In this work, we present a garnet-based SHO that operates within the positive nonlinear frequency shift regime while being magnetized in the film plane, enabling efficient spin-wave emission tunable in frequency with long-range propagation.
\begin{figure*}
    \centering
    \includegraphics{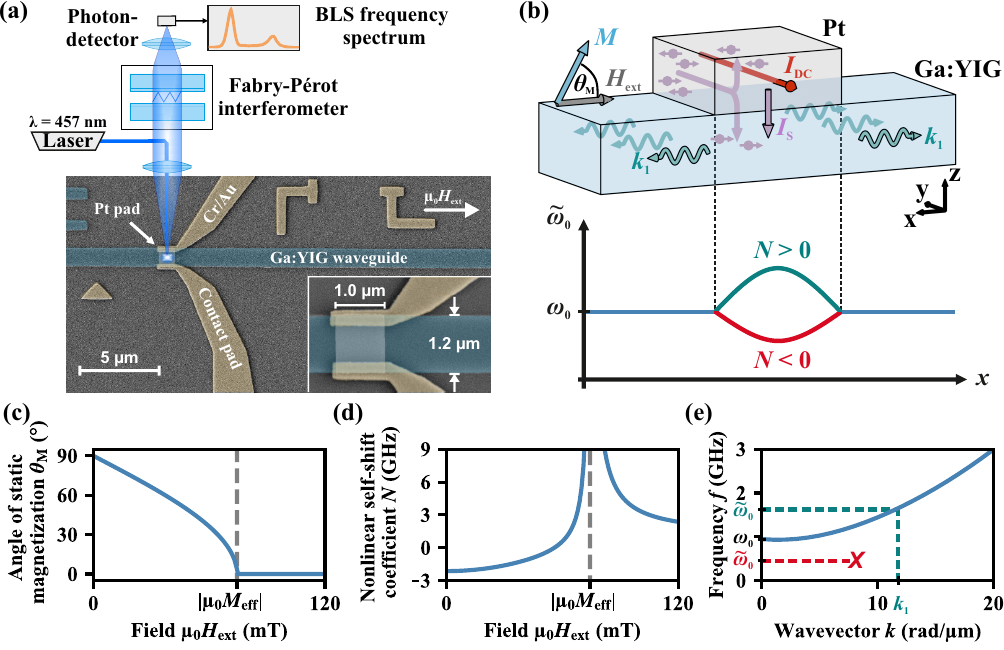}
    \caption{Experimental setup and Ga:YIG characteristics \textbf{(a)} Colorized SEM micrograph of the microstructure and a schematic of the applied BLS microscope. \textbf{(b)} Schematic illustration of the device layout with a frequency landscape for both nonlinear shift directions below. \textbf{(c)} Angle of the static magnetization $\theta_\mathrm{M}$ as a function of the external in-plane field $\upmu_0H_\mathrm{ext}$. \textbf{(d)} Nonlinear self-shift coefficient $N$ of the FMR mode as a function of $\upmu_0H_\mathrm{ext}$. \textbf{(e)}
    Calculated dispersion relation $f(k)$ for a Ga:YIG film in the linear case at $\upmu_0H_\mathrm{ext} =$ \SI{87}{\milli\tesla}, with two exemplarily included nonlinear auto-oscillation frequencies $\tilde{\omega_0}$.
    The calculations from (c)-(e) are presented in the Supplementary material, Sec. 1 and the material parameters are taken from Sec. 2.}
    \label{fig:Figure-1}
\end{figure*}
This is achieved by using a gallium-substituted yttrium-iron-garnet thin film\cite{dubs.gayig.2025} (Ga:YIG), which exhibits uniaxial anisotropy $H_\mathrm{U}$ along the film normal that is larger than the saturation magnetization $M_\mathrm{S}$, resulting in a dominant PMA\cite{carmiggelt.gayig.2021,böttcher.gayig.2022,breitbach.erasing.2024,breitbach.neuron.2025}. This is characterized by a negative effective magnetization $M_\mathrm{eff} = M_\mathrm{S}-H_\mathrm{U} <0$, leading to a positive nonlinear frequency shift for systems saturated in-plane by external fields, which represents the key distinction from pure YIG in this study. 
Consequently, the SHO based on in-plane magnetized Ga:YIG can operate at maximum SHE efficiency while being within the positive nonlinear shift regime to overcome the self-localized excitation.
Our measurements confirm the emission of spin waves originating from the confined auto-oscillation region with overall tunability in frequency of approximately \SI{1.6}{\giga\hertz} by varying the direct charge current, which corresponds to a relative change of more than \SI{150}{\percent}. In our study, we measure the excited spin waves up to a distance of more than \SI{10}{\micro\meter}, demonstrating that Ga:YIG is a potential candidate for efficient long-range coupling of multiple SHOs via spin waves.

\section{Characteristics of Ga:YIG and the microstructure}\label{sec:2}
Figure \ref{fig:Figure-1} (a) depicts the experimental setup and the fabricated microstructure under study. 
The sample is studied via space-resolved Brillouin light scattering (BLS) spectroscopy\cite{sebastian.BLS.2015}, which utilizes a laser ($\lambda = \SI{457}{\nano\meter}$) with a power of $P=\SI{1.5}{\milli\watt}$ on the sample. The laser is focused on the Ga:YIG surface through the substrate, providing access to the magnetization dynamics below metallic structures. 
The microstructure is based on a Ga:YIG waveguide with a thickness of \SI{55}{\nano\meter}, which was patterned by ion beam etching from a full film grown by liquid phase epitaxy on an (111)-oriented gadolinium gallium garnet substrate \cite{dubs.LPE.2017,dubs.LPE.2020}. A gallium substitution content of $x\approx1$ in the underlying Ga:YIG ($\mathrm{Y}_\mathrm{3}\mathrm{Fe}_\mathrm{5-x}\mathrm{Ga}_\mathrm{x}\mathrm{O}_\mathrm{12}$) results in a reduction of the saturation magnetization to $\upmu_{\mathrm{0}}M_\mathrm{S} = \SI{20.38}{\milli\tesla}$, which in combination with a relative film-substrate lattice mismatch induces robust PMA\cite{dubs.gayig.2025}. Note that nm-thin substituted YIG films, such as Bi:YIG, have also been observed to exhibit strong PMA, which classifies these films as suitable candidates as well. However, it has been reported that the robustness of PMA decreases significantly with increasing film thickness and, in comparison to Ga:YIG, larger frequency linewidths are often observed, based on a benchmark provided by Dubs \textit{et al.}\cite{dubs.gayig.2025}. Consequently, utilizing Bi:YIG in this study would require a film of a reduced thickness compared to the studied Ga:YIG to maintain robust PMA, which may also be accompanied by enhanced spin-wave damping, with both contributing to a reduction in spin-wave propagation lengths. Furthermore, a notable distinction between these material systems arises from the lattice site substituted by Ga, in contrast to substitutions with Bi, Mn, Tm, or other elements. In particular, Ga substitutes for Fe in YIG, while the other elements substitute for Y, resulting in a tunable $M_\mathrm{S}$ exclusively in Ga:YIG. This enables a reduced $M_\mathrm{S}$, which in turn decreases the auto-oscillation threshold \cite{ralph.STT.2008}.\\
The waveguide exhibits a width of \SI{1.2}{\micro\meter}, which was selected to minimize the complexity of the fabrication process for this study. A confined Pt pad is deposited on the waveguide with a thickness of \SI{6.5}{\nano\meter} and a width of \SI{1}{\micro\meter}, to which a current can be applied via contacts composed of Cr/Au (\SI{10}{\nano\meter}/\SI{100}{\nano\meter}). 
As shown in Fig. \ref{fig:Figure-1} (b), the applied charge current $I_\mathrm{DC}$ injects a spin current $I_\mathrm{S}$ via SHE into Ga:YIG, which induces SOTs that effectively counteract and compensate the spin-wave damping, resulting in a coherent auto-oscillation when a critical current is applied\cite{slavin.STNO.2009}.
While the emission of propagating spin waves is the objective of this device, the radiation of these waves also establishes an additional loss channel, which is expected to increase the auto-oscillation threshold. By confining the system to one dimension using a waveguide, these radiation losses can be effectively reduced, and the emitted spin-wave signal is guided, demonstrating the feasibility for future magnonic circuits.\\
The auto-oscillation is generally observed to occur at the minimum frequency of the local dispersion relation\cite{demidov.BEC-Spincurrent.2017}, which corresponds in Ga:YIG to the ferromagnetic resonance (FMR) at $k=0$ \cite{breitbach.erasing.2024}.
For large spin-wave amplitudes, the nonlinear nature of magnetization dynamics emerges, including the nonlinear frequency shift\cite{krivosik.hamiltonian.2010}:
\begin{equation}\label{Eq:Nonlinear-Shift}
    \tilde{\omega_0} = \omega_0+N\cdot|c_0|^2.
\end{equation}
Here, it can be defined by the nonlinear self-shift coefficient $N$ of the FMR mode and its spin-wave intensity $|c_0|^2$, nonlinearly shifting the auto-oscillation frequency $\tilde{\omega_0}$ compared to the linear FMR frequency $\omega_\mathrm{0}$. Therefore, depending on the sign of $N$, $\tilde{\omega_0}$ is shifted locally above or below $\omega_\mathrm{0}$, which remains constant in the surrounding medium, as illustrated below.
\begin{figure}
    \centering
    \includegraphics{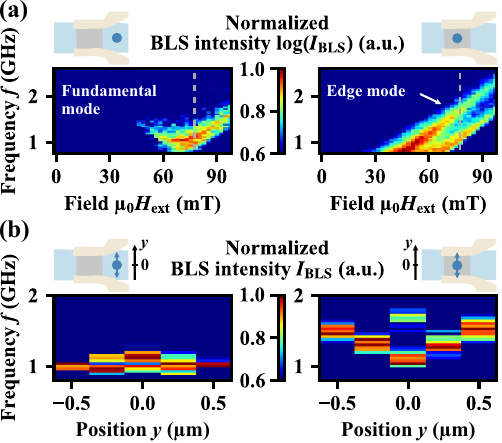}
    \caption{Thermal mode spectrum \textbf{(a)} Thermal BLS spectrum as a function of $\upmu_\mathrm{0}H_\mathrm{ext}$, measured on the bare Ga:YIG waveguide (left) and on Ga:YIG/Pt (right). The gray dashed line represents the field value $\upmu_\mathrm{0}H_\mathrm{app} = \SI{77.5}{\milli\tesla}$. \textbf{(b)} Thermal BLS spectrum on Ga:YIG (left) and Ga:YIG/Pt (right) at $\upmu_\mathrm{0}H_\mathrm{app}$ as a function of the measurement position $y$. The BLS spectrum of each position is normalized to the respective maximum.}
    \label{fig:Figure-2}
\end{figure}
Figure \ref{fig:Figure-1} (c) depicts the calculated angle of static magnetization $\theta_\mathrm{M}$ in Ga:YIG as a function of the in-plane external field $\upmu_\mathrm{0}H_\mathrm{ext}$. The magnetization initially points out-of-plane due to the PMA and is dragged gradually in-plane by increasing field, aligning with the external field for $H_\mathrm{ext}>|M_\mathrm{eff}|$.
Figure \ref{fig:Figure-1} (d) depicts the calculation of $N$ as a function of $\upmu_\mathrm{0}H_\mathrm{ext}$, revealing  a transition from negative to positive shifts for increasing fields, being maximized at the compensation field $H_\mathrm{ext} = |M_\mathrm{eff}|$. The divergence observed in this configuration can be attributed to the compensation of PMA by the external field, resulting in the elimination of an energetically favored orientation. 
We select the external field value for the operation of the SHO slightly above |$\upmu_0M_\mathrm{eff}$| in order to ensure maximum SHE efficiency with a large positive nonlinear shift coefficient $N$. 
However, a certain separation to $|M_\mathrm{eff}|$ is necessary, as $\omega_0$ approaches zero for fields close to $|M_\mathrm{eff}|$ (see Supplementary material, Sec. 1), which can be attributed to the formation of domains\cite{Grassi.Domain.2022}. Moreover, the utilized BLS setup detects frequencies only above \SI{0.6}{\giga\hertz}. Therefore, a field value of approximately \SI{12.5}{\milli\tesla} larger than $|\upmu_0M_\mathrm{eff}|$ is selected, resulting in a FMR frequency of around \SI{1}{\giga\hertz} and a well-defined in-plane magnetization.
The calculated linear dispersion relation of Ga:YIG is depicted in Fig. \ref{fig:Figure-1} (e) for $\upmu_\mathrm{0}H_\mathrm{ext} =$ \SI{87}{\milli\tesla}, which is characterized by a FMR frequency of around \SI{1}{\giga\hertz} and a parabolic shape, which aligns with previous studies\cite{carmiggelt.gayig.2021,breitbach.erasing.2024}. Note that $|\upmu_0M_\mathrm{eff}|$ of the unprocessed Ga:YIG thin film differs from the one of the waveguide, resulting in a different selected external field for Fig. \ref{fig:Figure-1} (e), than in the following measurements. Furthermore, nonlinearly shifted auto-oscillation frequencies $\tilde{\omega_0}$ are exemplarily included to visualize the concept of spin-wave emission. 
A negatively shifted auto-oscillation frequency ($N<0$) has no corresponding spin-wave state outside the injector, suppressing spin-wave emission. A positive shift ($N>0$), however, results in an auto-oscillation frequency, which corresponds to a propagating mode with wavevector $k_\mathrm{1}$ in the surrounding material, enabling spin-wave emission.\\
FMR spectroscopy on reference samples (see Supplementary material, Sec. 2) reveals a reduction of $\upmu_0H_\mathrm{U}$ by around \SI{3}{\milli\tesla} and a damping enhancement by almost fivefold due to the Pt layer, effects analogous observed in YIG/Pt \cite{sun.YIG/Pt-damping.2013,beaulieu.anisotropyYIG.2018}. As the damping is only enhanced in the SHO region covered by Pt, propagation in the bare waveguide remains at low damping  ($\alpha_\mathrm{Ga:YIG} \approx$ \SI{4e-4}{}). Furthermore, the damping in Ga:YIG/Pt remains an order of magnitude below that observed in metal-based SHOs\cite{zahedinejad.cmos-SHNO.2018}.\\
Figure \ref{fig:Figure-2} depicts a spatial BLS analysis of the microstructure, showing the spin-wave population as a function of frequency in the absence of excitation or driving currents, which is referred to as the thermal BLS spectrum.
Two spectra, obtained on the bare Ga:YIG waveguide (left graph) and on Ga:YIG/Pt (right graph), are shown in Fig. \ref{fig:Figure-2} (a) as a function of $\upmu_\mathrm{0}H_\mathrm{ext}$.
The BLS signal vanishes for both at low fields, as the measurement technique is mainly sensitive to the dynamic out-of-plane magnetization. The minimum frequency occurs at the compensation field $H_\mathrm{comp}$, allowing for an approximation of $M_\mathrm{eff}$ (see Supplementary material, Sec. 2).
From the detected frequencies on the bare Ga:YIG waveguide, we conclude a local $\upmu_0M_\mathrm{eff} \approx $ \SI{-65}{\milli\tesla}, which corresponds to $\upmu_0H_\mathrm{U,Ga:YIG} \approx$ \SI{85}{\milli\tesla}, when $M_\mathrm{S}$ is assumed to be unchanged by fabrication.
The Ga:YIG/Pt region, however, reveals clear modifications, particularly the presence of two distinct modes. While the mode at lower frequency aligns approximately with the mode on bare Ga:YIG, the origin of the additional one appears to be a result of the microstructuring, as the Ga:YIG/Pt full-film reference sample exhibits only a single mode.
The thermal BLS spectrum for five positions across the waveguide width is depicted in Fig. \ref{fig:Figure-2} (b) to obtain a spatial profile of these modes. While the \textit{fundamental mode} remains constant across the width on the bare Ga:YIG waveguide, it appears to increase in frequency towards the edges on Ga:YIG/Pt.
In the outer position, the frequency aligns approximately with the frequency of the additional mode, which is only observed in the center simultaneously with the fundamental mode.
The occurrence of this mode can be attributed to a reduction in uniaxial anisotropy toward the edges, which is why the second mode will be referred to as \textit{edge mode}. This effect arises from the contact structures located at the waveguide edges, which generate enhanced strain. The local reduction of uniaxial anisotropy due to Pt and Au structures has already been demonstrated on Bi:YIG and YIG, respectively \cite{evelt.biyig.2018,Wu.ANisotropyYIG.2020}. The uniaxial anisotropy of the edge mode can be estimated to $\upmu_0H_\mathrm{U,edge} \approx$ \SI{65}{\milli\tesla}, which is confirmed via micromagnetic simulations shown in Sec. \ref{sec:5}.
For the following experiments an external field of $\upmu_\mathrm{0}H_\mathrm{app} = $ \SI{77.5}{\milli\tesla} is applied to ensure that the waveguide is in-plane magnetized and exhibits a FMR frequency of approximately \SI{1}{\giga\hertz}. In conclusion, it is demonstrated that the mode structure of Ga:YIG/Pt can be modified through the fabrication of microstructures, revealing a richer mode spectrum. The local modification of material parameters has been identified as a method for enabling spin-wave emission \cite{demidov.excitation.2016,divinskiy.sw-emission.2018}. While these approaches rely on the variation of magnetic material geometry, the underlying microstructure exhibits a local change of magnetic properties. However, the present study focuses on the utilization of the nonlinear positive frequency shift for tunable emission of spin waves, independent of this effect.
\section{Local auto-oscillation}
\begin{figure}
    \centering
    \includegraphics{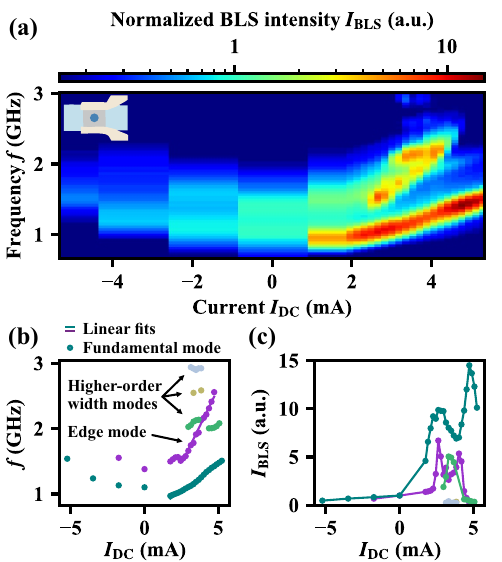}
    \caption{Auto-oscillation \textbf{(a)} BLS spectrum, obtained through averaging across the Pt pad region, as a function of the current $I_\mathrm{DC}$. The data is normalized to the thermal intensity maximum at $I_\mathrm{DC}=\SI{0}{\milli\ampere}$. The spectral intensity maxima for each current value were fitted by a Gaussian, with the extracted frequency and intensity depicted as a function of $I_\mathrm{DC}$ in \textbf{(b)} and \textbf{(c)}, respectively.}
    \label{fig:Figure-3}
\end{figure}

To investigate the auto-oscillation properties of the SHO, the BLS spectrum on Ga:YIG/Pt is shown in Figure \ref{fig:Figure-3} (a) as a function of the applied current $I_\mathrm{DC}$. The SHE-induced control via the current is demonstrated by the significant intensity asymmetry, as the current direction switches between damping enhancement and compensation. Both thermal modes appear to be populated for increasing currents and a positive frequency shift is observed.\\
In Fig. \ref{fig:Figure-3} (b), the frequency as a function of $I_\mathrm{DC}$ is shown for the different modes observed. While the fundamental- and edge-mode frequencies exhibit a linear current dependence, three additional modes emerge within a certain current range, which remain approximately at a constant frequency. 
These can be associated with higher-order width modes, following theoretical calculations\cite{koerber.tetrax.2021,koerber.tetraX.2022} of the Ga:YIG waveguide (see Supplementary material, Sec. 3).
However, the fundamental and edge mode are of specific interest, given their tunability in frequency and significant intensity, effectively representing two coupled SHOs within one spin-current injection region. They exhibit a different slope in the current-induced frequency shift, with an almost three times larger slope of the edge mode compared to the fundamental mode.
While the edge mode can be tuned from approximately \SI{1.4}{\giga\hertz} to \SI{2.6}{\giga\hertz}, the fundamental mode covers the lower frequencies from \SI{1.0}{\giga\hertz} to \SI{1.5}{\giga\hertz}, leading to an overall bandwidth of \SI{1.6}{\giga\hertz}.
Also for the reversed current direction, where spin waves are strongly suppressed, a frequency shift is observed, suggesting that the frequency shift is not solely caused by the spin-wave intensities. A potential impact of heating must be considered, which will be further discussed on the basis of the micromagnetic simulations in Sec. \ref{sec:5}.\\
The extracted intensity $I_\mathrm{BLS}$ is depicted in Fig. \ref{fig:Figure-3} (c), exhibiting a slight intensity decrease for negative currents due to damping enhancement.
At positive currents, the threshold current for auto-oscillation of the fundamental mode is the lowest with about \SI{1}{\milli\ampere}, which corresponds to a current density of $j_\mathrm{DC}=$ \SI{1.54e11}{\ampere\per\meter\squared}. The edge mode has a threshold value that is approximately twice as high, which is presumably due to a locally increased damping. Note that the threshold current for each mode is estimated by linearly
extrapolating from the initial increase in intensity to the current of vanishing BLS intensity.
The fundamental mode increases to a maximum intensity level of about 15 times the thermal signal but shows a decline for intermediate currents, which is correlated with the multi-mode excitation, indicating mode competition.
For currents $I_\mathrm{DC} > \SI{5}{\milli\ampere}$, the fundamental mode dominates and is the only to be excited. However, its intensity exhibits a declining trend towards higher currents, which may result from heating, particularly given that the Curie temperature of Ga:YIG is only around \SI{400}{\kelvin}\cite{dubs.gayig.2025}. Consequently, higher currents that might overheat the system are not applied.
To conclude this part, a dynamic auto-oscillation frequency range of more than \SI{1}{\giga\hertz} corresponding to propagating modes in the adjacent waveguide is accessible.
\section{Spin-wave emission}
\begin{figure}
    \centering
    \includegraphics{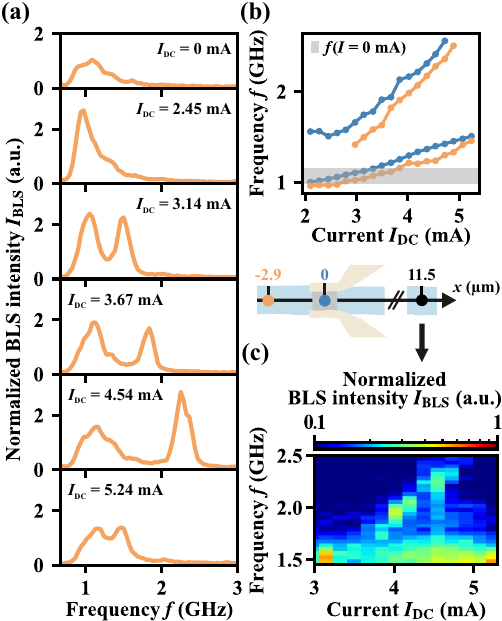}
    \caption{Spin-wave emission \textbf{(a)} BLS spectra obtained at $x = \SI{-2.89}{\micro\meter}$ for several currents $I_\mathrm{DC}$, normalized to the maximum value without a current applied. \textbf{(b)} Extracted frequencies of the edge and fundamental mode as a function of $I_\mathrm{DC}$ for two measurement positions (indicated in the pictogram). \textbf{(c)} BLS spectrum of the edge mode as a function of $I_\mathrm{DC}$ that is obtained in a distance of \SI{11.5}{\micro\meter} to the Pt pad.}
    \label{fig:Figure-4}
\end{figure}

To investigate spin-wave emission, BLS measurements are performed at various distances from the Pt pad, as indicated by the colored dots in the schematic of Fig. \ref{fig:Figure-4}.
Figure \ref{fig:Figure-4} (a) illustrates the BLS spectra in a distance of $x =$ \SI{-2.9}{\micro\meter} for several currents, including the thermal spin-wave spectrum of bare Ga:YIG for $I_\mathrm{DC} =\SI{0}{\milli\ampere}$, with a distribution around \SI{1}{\giga\hertz}. The other BLS spectra exhibit two intensity maxima that exceed the thermal reference, clearly demonstrating the emission of spin waves from the SHO. 
In the waveguide, the ratio of the intensities of the two propagating modes differs from the ratio on the Pt pad. The high-frequency mode exhibits an intensity level that is comparable to the low-frequency mode, despite the latter having a substantial overlap with the thermal spectrum. This change in ratio can be attributed to a higher group velocity of the high-frequency mode, which is expected due to the parabolic dispersion relation (see Fig. \ref{fig:Figure-1} (e)).
\begin{figure*}
    \centering
    \includegraphics{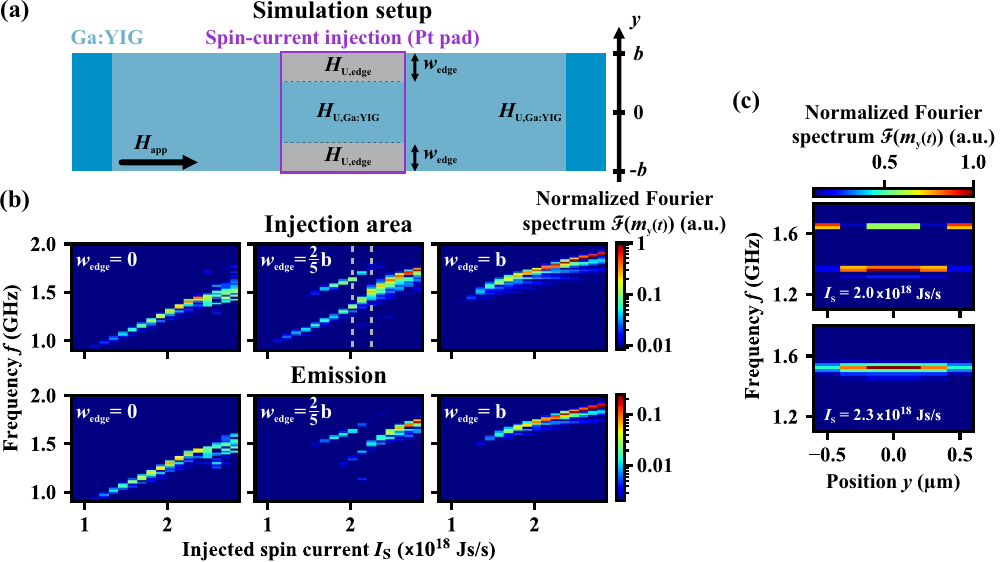}
    \caption{Micromagnetic study \textbf{(a)} Schematic illustration of the simulation setup. Towards the waveguide ends (darker blue region), the damping is smoothly increased to avoid reflections and mimic an infinite system. More detailed simulation parameters are given in the Supplementary material, Sec. 5A. \textbf{(b)} Normalized Fourier spectra $\mathcal{F}(m_y(t))$ as a function of the injected spin current $I_\mathrm{S}$, obtained on the injector and in a distance of \SI{1.5}{\micro\meter} for three edge widths $w_\mathrm{edge}$. 
    All of them are normalized to the maximum value observed within the entire simulation series. The gray dashed lines represent the current density values used for (c). \textbf{(c)} $\mathcal{F}(m_y(t))$ as a function of the position $y$ on the injector for two injected spin currents.}
    \label{fig:Figure-5}
\end{figure*}
Note that only two propagating modes can be observed, indicating that the excited higher-order width modes are either emitted less efficiently, have lower propagation length or are detected poorly by BLS.\\ 
The frequency maxima of the two propagating modes are depicted in Fig. \ref{fig:Figure-4} (b) as a function of $I_\mathrm{DC}$, with the auto-oscillation data from Fig. \ref{fig:Figure-3} (b) included.
The emission frequency is found to be reciprocal (see Supplementary material, Sec. 4).
The current dependence of the propagating and auto-oscillation frequencies appears similar, confirming that the propagating spin-wave modes originate from the auto-oscillation of the fundamental and edge modes. The observed discrepancy of approximately \SI{0.2}{\giga\hertz} at a specific current value can be attributed to the local heating by the laser spot during the measurement on Ga:YIG/Pt. The observed frequencies of the edge mode correspond to wavevectors from $k=\SI{10}{\radian\per\micro\meter}$ up to $k=\SI{17}{\radian\per\micro\meter}$ according to the dispersion relation in Fig. \ref{fig:Figure-1} (e).\\
The emitted signal from the edge mode propagates along the Ga:YIG waveguide and can still be detected in a distance of \SI{11.5}{\micro\meter}, which is shown in Fig. \ref{fig:Figure-4} (c), demonstrating the potential of Ga:YIG to enable an efficient long-range interaction in a network of SHOs.
\section{Micromagnetic study}\label{sec:5}
To verify our interpretations, we study the underlying system using micromagnetic simulations\cite{vansteenkiste.mumax.2014, Aithericon}. Figure \ref{fig:Figure-5} (a) provides a schematic illustration of the simulation setup consisting of a waveguide with a confined spin-current injector, including two edge regions with reduced out-of-plane anisotropy.
Figure \ref{fig:Figure-5} (b) depicts the normalized Fourier spectra of the dynamic magnetization $\mathcal{F}(m_y(t))$ obtained on the injector and in a distance of \SI{1.5}{\micro\meter} as a function of the injected spin current $I_\mathrm{S}$ for different edge widths $w_\mathrm{edge}$. 
Note that a current density $j_\mathrm{DC}$ is applied in the micromagnetic simulation. However, the impact of this current density is not directly comparable to the experimental study, as the actual efficiency of charge-to-spin current conversion in the underlying microstructure remains unknown due to undetermined parameters, such as the spin-Hall angle of Pt or the spin-mixing conductance of the Ga:YIG/Pt interface. Therefore, the injected spin current $I_\mathrm{S}$ can only be determined in the micromagnetic study (see Supplementary material, Sec. 5B), which restricts the comparison of experiment and simulation to a qualitative level. It can be observed that a homogeneous pad with either $H_\mathrm{U,Ga:YIG}$ ($w_\mathrm{edge}=0$) or $H_\mathrm{U,edge}$ ($w_\mathrm{edge}=b$) generates a single-mode auto-oscillation and emission, which exhibits either a frequency bandwidth from \SI{1.0}{\giga\hertz} to \SI{1.5}{\giga\hertz} or from \SI{1.5}{\giga\hertz} to \SI{2.0}{\giga\hertz}, respectively. 
The case featuring edge regions is illustrated in the center, whose edge width of $w_\mathrm{edge}=\frac{2}{5}b$ corresponds to the actual size of the gold contacts on the microstructure. Here, both individual modes coexist within a specific spin-current range, with the fundamental mode being dominant elsewhere. At intermediate spin currents, the edge mode is more pronounced in emission, while the fundamental mode becomes dominant at higher spin currents. This agrees well with the experiment and confirms that the two-mode system originates from edge regions of lower PMA, effectively creating distinct oscillators on the injector. However, the ratio between auto-oscillation and emission of the fundamental mode is less pronounced for the two-mode system than it is for single-mode, revealing a certain interaction of both oscillators. This might be either due to an induced nonlinear frequency (cross-)shift from the edge mode to the fundamental mode, which has not been considered in Sec. \ref{sec:2}, or simply because of a smaller effective oscillator size.\\
As the frequency shift of the fundamental mode aligns for both simulation and experiment, mere heat does not have a significant impact within the mainly used spin-current regime. However, the frequency shift of the edge mode is similar to that of the fundamental mode in the simulation, which contradicts to the experimentally observed three-fold enhancement. This effect can be attributed to an additional strain induced via the thermal expansion of the Pt pad or gold contacts, to which the edge mode is more sensitive.
This phenomenon effectively extends the frequency bandwidth of the SHO within the accessible current range. 
Note that the experimentally observed frequency shift for negative currents is not reproduced by simulations (see Supplementary material, Sec. 5C), which aligns with Eq. \ref{Eq:Nonlinear-Shift} but contradicts with the experiment.\\
Figure \ref{fig:Figure-5} (c) depicts the normalized Fourier spectrum $\mathcal{F}(m_y(t))$ as a function of the position across the width on the Pt pad for two injected spin currents, which represent the two- and single-mode domain. 
The two-mode profile aligns with Fig. \ref{fig:Figure-2} (b), with the edge mode being present at the edges and in the center of the Pt Pad, while the fundamental mode is situated only in the center.
The transition to the single-mode system is observed when the frequency of the fundamental mode approaches the initial frequency level of the edge region, resulting in a single coherent auto-oscillation across the entire width, as evidenced by its width profile in Fig. \ref{fig:Figure-5} (c). This suggests that a single-mode system emerges whenever the spatial overlap between the fundamental and edge mode can be established, which corresponds to an increased cross-mode coupling according to a recent theoretical model \cite{verba.STNO-multi-mode.2025}.
\section{Conclusion}
In conclusion, we demonstrated a garnet-based SHO that utilizes low-damping Ga:YIG with robust PMA to obtain an operation regime, which provides large SHE efficiency while also enabling resonant spin-wave emission. 
Current-control allows to tune the operation frequency over a continuous bandwidth of \SI{1}{\giga\hertz}, while the low spin-wave damping of the system provides a significant propagation length of at least \SI{10}{\micro\meter}. We revealed different regimes of operation, ranging from single- to multi-mode excitation, extending the full SHO bandwidth to approximately \SI{1.6}{\giga\hertz}. Our studies indicate that the emergence of an edge region with reduced PMA due to microstructuring is responsible for the two-mode behavior, which we confirm using micromagnetic simulations. These insights enable future experimental designs to accurately control the SHO mode structure by reducing structure dimensions or by adjustments in the fabrication process. Ultimately, the implementation of spin-wave sources based on low-damping magnetic garnets with PMA represents a promising candidate for a building block in future magnonic circuits.

\section*{supplementary material}
The supplementary material contains additional information on the calculation of the equilibrium magnetization $\theta_\mathrm{M}$, the nonlinear self-coefficient $N$ of the FMR mode and the FMR frequency as a function of the external field $\omega_0(H_\mathrm{ext})$, on the FMR spectroscopy study of full-film reference samples, on the simulated dispersion relation, on the reciprocal emission character of the SHO, and on the micromagnetic simulations.

\begin{acknowledgements}
This research was funded by the European Research Council within the Starting Grant No. 101042439 "CoSpiN" and by the Deutsche Forschungsgemeinschaft (DFG, German Research Foundation) within the Transregional Collaborative Research Center—TRR 173–268565370 “Spin + X” (project B01), as well as by the DFG project - 271741898.
\end{acknowledgements}

\section*{Author declarations}
\subsection*{Conflict of Interest}
The authors have no conflicts to disclose.

\section*{Data availability}
The data that support the findings of this study are available from the corresponding author upon reasonable request.

\section*{References}

%

\end{document}